\documentclass{article}
\usepackage{spconf,amsmath,graphicx}
\usepackage{multirow}
\usepackage{bm}
\usepackage{marvosym}
\title{AISPEECH-SJTU accent identification system for the Accented English Speech Recognition Challenge}
\name{$^{\dagger}$Houjun Huang$^{1}$$^{,2}$, $^{\dagger}$Xu Xiang$^{1}$, Yexin Yang$^{2}$, Rao Ma$^{2}$, \textsuperscript{\Letter}Yanmin Qian$^{2}$ \thanks{ ${\dagger}:$These authors contributed equally to this work. Yanmin Qian is the corresponding author. This work was supported by the National Key R\&D Program of China (No. 2018YFB1004602) and the China NSFC project (No. 62071288)}}

\address{
 $^{1}$AISpeech Ltd, Suzhou China\\
 $^{2}$MoE Key Lab of Artificial Intelligence, AI Institute SpeechLab, Department of Computer Science\\and Engineering Shanghai Jiao Tong University, Shanghai,China \\
 \{houjun.huang, xu.xiang\}@aispeech.com, \{yangyexin, rm1031, yanminqian\}@sjtu.edu.cn
 }

\newcommand{\etal}{\textit{et al.}}
 
\begin{document}
\ninept
\maketitle

\begin{abstract}
This paper describes the AISpeech-SJTU system for the accent identification track of the Interspeech-2020 Accented English Speech Recognition Challenge.
In this challenge track, only 160-hour accented English data collected from 8 countries and the auxiliary Librispeech dataset are provided for training.
To build an accurate and robust accent identification system, we explore the whole system pipeline in detail.
First, we introduce the ASR based phone posteriorgram (PPG) feature to accent identification and verify its efficacy.
Then, a novel TTS based approach is carefully designed to augment the very limited accent training data for the first time.
Finally, we propose the test time augmentation and embedding fusion schemes to further improve the system performance.
Our final system is ranked first in the challenge and outperforms all the other participants by a large margin.
The submitted system achieves 83.63\% average accuracy on the challenge evaluation data, ahead of the others by more than 10\% in absolute terms.

\end{abstract}
\section{Introduction}
\label{sec:intro}
An accent is a manner of pronunciation peculiar to a particular individual, location, or nation.
It may be influenced by the speaker's locality, education attainment or first language.
For the pervasiveness of accents, accent identification is widely utilized in robust speech recognition, speaker recognition, language identification and forensic applications.

Earlier studies in accent identification mainly focus on combining linguistic theory with statistical analysis. Piat~\etal, used a statistical approach based on prosodic parameters and found that the duration and energy are promising parameters for correct identification~\cite{piat2008foreign}. Berkling~\etal, leveraged the structure of English syllable to improve the accent identification~\cite{berkling1998improving}. Chen~\etal, proposed a Gaussian mixture model (GMM) based method for Mandarin accent identification~\cite{chen2001automatic}. Recently, deep neural network based approaches have emerged in this field. Weninger~\etal, made use of bidirectional Long Short-Term Memory (bLSTM) networks to model longer-term acoustic context~\cite{weninger2019deep}. Jiao~\etal, proposed a system utilizing long and short term features in parallel using DNNs and RNNs~\cite{jiao2016accent}.

In this work, we describe our accent identification system for the Interspeech-2020 Accented English Speech Recognition Challenge  (AESRC)~\cite{AESRC2020} in detail.
Since accent training data is rather limited, it is critical to effectively make use of the auxiliary Librispeech dataset.
In contrast to the previous studies, we have three distinct contributions.
First, we introduce the ASR based PPGs as the discriminative features for accent identification.
Second, we propose a novel TTS based approach to synthesize the accent data, which provides richer speaker, channel and text variability for training the accent classifier.
Third, we develop the test time augmentation and the hierarchical multi-embedding joint model for improving system performance.

This paper is arranged as follows.
Section~\ref{sec:sys} gives an in-depth description of the framework of our system.
Section~\ref{sec:exp} presents our experiments with different settings.
Section~\ref{sec:conclusion} concludes our paper.

\section{System Description}
\label{sec:sys}
\begin{figure}[!th]
    \centering
    \includegraphics[scale=0.45]{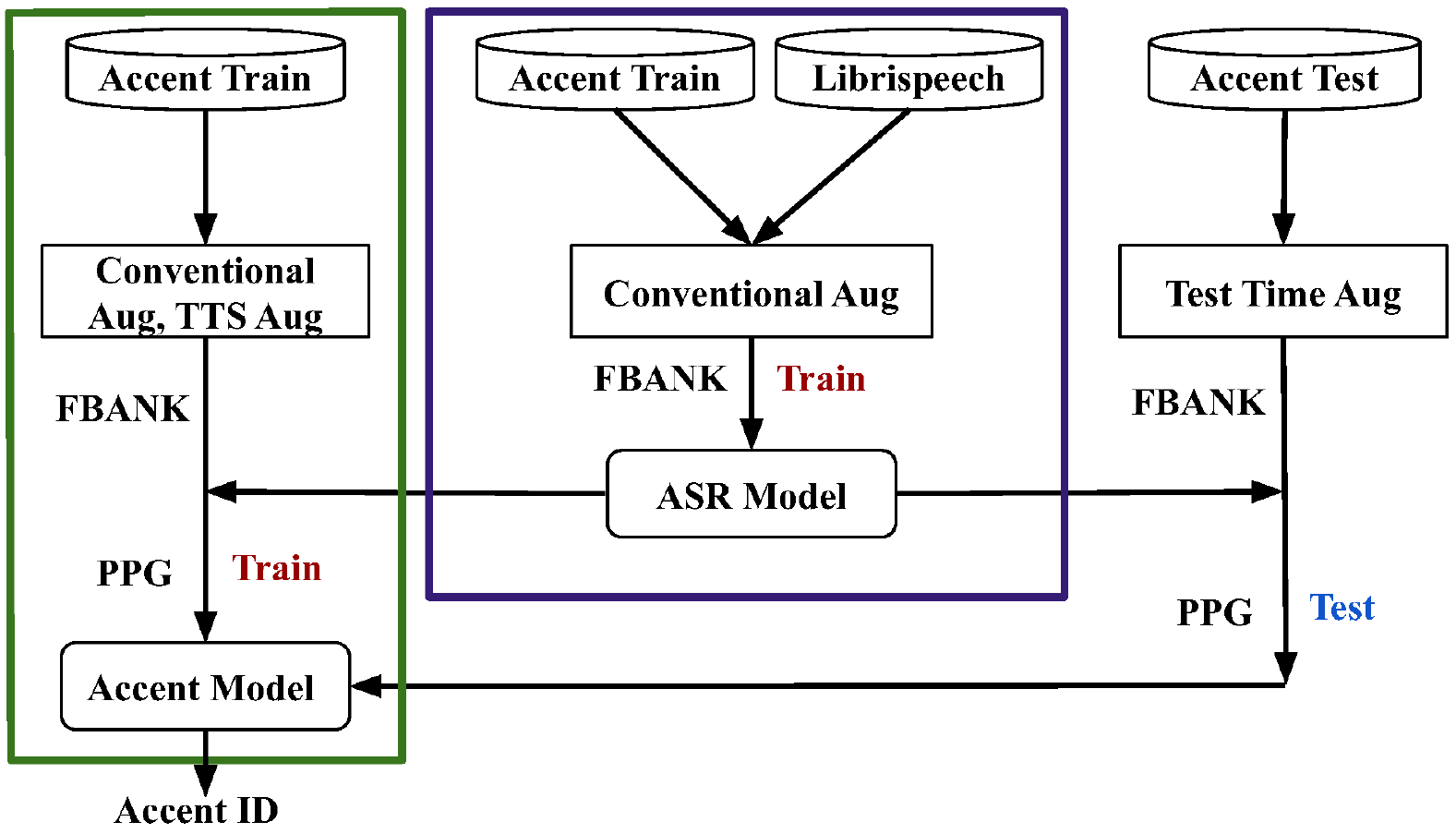}
    \caption{Our system diagram for the challenge. Firstly, an ASR model is trained on the pooled data. Then, the PPG features derived from the ASR model are prepared to train an AID model. Finally, The AID model makes predictions on the test data using PPG features.}
    \label{fig:my_label}
\end{figure}

In this section, we depict our system for the accent identification challenge, which is shown in Figure~\ref{fig:my_label}.

First, we leverage both the accent training data and the Librispeech data to train an ASR model with conventional data augmentation. The PPG features are extracted from the ASR model and employed to train the accent identification(AID) model.
Then, we propose a novel TTS based data augmentation to augment the accent data for training an accurate and robust accent classifier.
Finally, we introduce the test time augmentation scheme to improve system performance on the test data.
Moreover, we further boost the system performance with the hierarchical multi-embedding joint model.

\subsection{ASR based PPG feature extraction}
\label{ppg}
While MFCC and FBANK features are popular in speech related tasks, we wouldn't build our AID system on them directly as the following reasons.
First, AID system built on MFCC or FBANK features directly could not take advantage of data-sets without accent labels
Second, as these features are low level and not task-oriented, they may contain some nuisance attributes like speaker or text specific, which makes the learning of accent related representation harder, especially when the amount of the training data is limited.

To address these issues, we adopt the phone posteriorgram (PPG) feature that has been successfully applied for cross-lingual voice conversion~\cite{sun2016personalized} and cross-accent voice conversion ~\cite{zhao2019foreign} to train the accent classifier.
PPG is a time-versus-class vector that represents the posterior probabilities of phonetic classes for a specific time frame.
In this work, we first train a speaker independent (SI) automatic speech recognition (ASR) model with both the accent training data and the Librispeech data and then extract the PPG features with the SI-ASR model.
With this process, the resulting PPG features have the speaker independent property which helps improve the robustness of the system.

To train a robust ASR model for PPG feature extraction, we employ two kinds of conventional data augmentation that have been widely used in automatic speech recognition tasks.
The first one augments original data with additive noise and reverberation.
For additive noise, the music, noise and speech part in the MUSAN dataset \cite{snyder2015musan} are used.
For reverberation, the room impulse responses (RIRs) and the simulated RIRs described in Kaldi's~\cite{povey2011kaldi} VoxCeleb recipe are used.
The second one is based on warping the signal in the time domain. We randomly change the tempo of the audio signal while ensuring that the pitch and spectral envelope of the signal unchanged. The {\it tempo} effect of the SoX tool was used to achieve such speech rate perturbation.

\subsection{TTS based data augmentation}
\label{tts-train}
Since only 160 hours of accent data are provided, it is too limited to train an accurate and robust accent identification model.
Therefore, we develop a novel TTS based data augmentation approach that is specially designed for synthesizing accent data.

Using the generated high-quality artificial speech as the augmented data has been successfully facilitated in ASR systems~\cite{sun2020generating}.
Recent advances in speech synthesis (text-to-speech, TTS) allow unsupervised modeling of prosody and speaker variations, which give the power to synthesize the same texts with diverse speaking styles. Moreover, the development of TTS models has made synthesized speech indistinguishable from human speech. 
% Considering only 160 hours of accent data with 20 hours for each accent are allowed to use in this challenge, it is too limited to train an accurate and robust accent classifier.
% To overcome the data sparsity issue, we develop a novel TTS based data augmentation approach to synthesize more accent data.
% To improve the accuracy of accent identification, we augment training data with speech synthesis. 
%We train 8 individual TTS models for each accent and synthesize speech data with each model for all the given speakers.
In this work, we choose FastSpeech~\cite{ren2019fastspeech} as our synthesizer and LPCNet~\cite{valin2019lpcnet} as the vocoder.
% In addition, we train a TDNN x-vector extractor on all available data for modeling speaker identities~\cite{snyder2018x}. 

% A 20-band Mel-spectrogram calculated with window of 20 ms and hop of 10 ms is used as the intermediate acoustic representation.

FastSpeech is a transformer-based model which generates the entire sequence in a non-autoregressive manner. The implementation of FastSpeech is based on the ESPnet toolkit~\cite{watanabe2018espnet}. We use the default settings as in~\cite{ren2019fastspeech}. Instead of applying the knowledge distillation process, we train the FastSpeech model from scratch only using the extracted features. The synthesizer converts input text to spectrogram. The size of the input vocabulary is 41, including English phonemes, a pause break token, and a sentence boundary token. Additionally, we augment the decoder with a five-layer post-net~\cite{shen2018natural}, which slightly enhance model performance. LPCNet is a variant of WaveRNN that combines linear prediction with recurrent neural networks, greatly promoting the synthesis efficiency~\cite{valin2019lpcnet}. Our implementation is based on~\cite{valin2019lpcnet}.
%, where sparsification is applied for lower time complexity.

% The TTS model training process is conducted as follows:
First, we train a TDNN x-vector speaker model~\cite{snyder2018x} with the pooled data consists of the accent training data and the auxiliary Librispeech data.
% Then for each utterance, its x-vector and the phoneme representations form a training sample for the FastSpeech synthesizer.
The x-vectors and the phoneme representation extracted from the pooled data are then used to train the FastSpeech synthesizer.
% For each training sample, the extracted x-vector is fed to the FastSpeech model together with the phoneme representations.
In order to better capture the characteristics of each accent, we create 8 accent specified synthesizer by finetuning the FastSpeech Model on each 20-hour accent data respectively. Finally, on the clean subset of the overall training data, we train two LPCNet vocoders for the male and female speakers respectively.

For each speaker from the accent training data, 30 utterances are grouped to calculate the speaker specific statistics.
We use these statistics with randomly selected reference texts to synthesize data with the previously trained accent specific synthesizers.
% every 30 utterances are grouped to calculate the speaker specific statistics and x-vector.
% We use the statistics and x-vector to synthesize speech with randomly select 40 reference texts.
With this process, the generated speech can preserve the speaker's speaking style while adopting new accents.
% The total size of the augmented dataset is approximately 11 times that of the original dataset. 

%\subsection{TDNN based accent identification model}
%
%
% note:
% embedding + projection layer
%
%
%In this challenge, we develop two models for accent identification: TDNN and RES2SETDNN.
%The architectures of TDNN and RES2SETDNN are shown in Figure~\ref{fig:tdnn}.
%The TDNN model is similar to the one used in~\cite{snyder2018x}, but is enlarged to 2$\times$ size, as we find it gives higher accuracy on the development data.
%The RES2SETDNN model is modified from the 

%\begin{figure}[!th]
%    \centering
%    \includegraphics[scale=0.4]{TDNN.pdf}
%    \caption{TDNN and RES2SETDNN model structures. We denote k for the kernel size and d for the dilation factor.}
%    \label{fig:tdnn}
%\end{figure}
%\textit{3*3} in SE-Res2Blocks(k=3,d=2) is \textit{Conv1D+ReLU+BN(k=3,d=2)}}

\subsection{Hierarchical multi-embedding joint model}
\label{hm}
\begin{figure}[!th]
    \centering
    \includegraphics[scale=0.45]{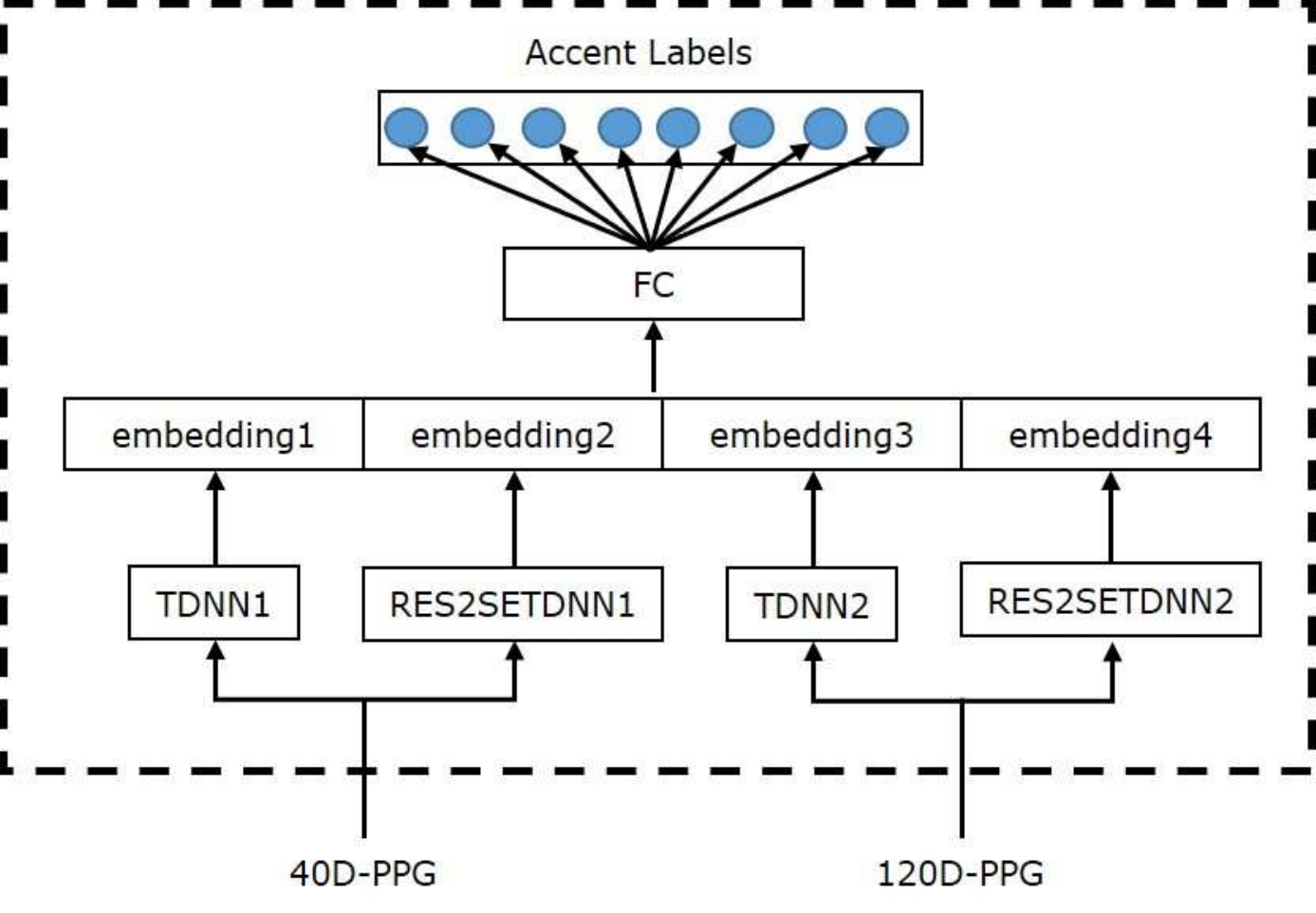}
    \caption{The hierarchical multi-embedding joint model includes the TDNN and RES2SETDNN sub-models and accepts two set of features. \textit{40D-PPG} denotes the 40-dimensional PPG features, \textit{120D-PPG} denotes the 40-dimensional PPG features and its first and second order difference and \textit{FC} denotes the fully connected linear layer.}
    \label{fig:jointtrain}
\end{figure}

% To improve system performance, we train multiple TDNN models and aggregate the discriminative power of each model by concatenating the embedding.
% Given the concatenated embedding, we train a 8-class DNN model on the accent training set, and then use it to make predictions on the test set.
In our final submission, we use a hierarchical multi-embedding joint model to predict the accent label, which is shown in Figure ~\ref{fig:jointtrain}. The structure of the TDNN sub-model is the same as the one in~\cite{snyder2018x}, but with 2$\times$ size, as we find it gives higher accuracy on the development data. The RES2SETDNN sub-model is developed in a way similar to~\cite{desplanques2020ecapa}, by introducing the Res2Net~\cite{gao2019res2net} type convolution and the squeeze-and-excitation~\cite{hu2018squeeze} module to the original TDNN structure. However, we do not include the residual connection in our model, for there is no improvement in performance.

We train the joint model in a progressive way as follows.
First, we pretrain each accent identification model independently using an additive angular margin softmax loss ~\cite{deng2019arcface,xiang2019margin}.
Then, for each model the embedding extraction part is fixed.
Finally, we train a linear regression classifier based on the concatenated embeddings extracted from each sub-model.

% Parameters of this model could be join trained directly or trained in a step method. In this work we train them by two steps:

% step1: Train the parameters of TDNN1, TDNN2, RES2SETDNN1 and RES2SETDNN2 with accent train data independently. Structure of TDNN based accent classifier and RES2SETDNN accent classifier is shown in Figure~\ref{fig:tdnn}

% step2: Train the parameters of the last full connection layer by fixing other parameters.

\begin{table*}[!thb]
  \caption{This table shows the identification accuracy (\%) of two model architectures on the development set: TDNN and RES2SETDNN. For each architecture, two kinds of feature FBANK and PPG and the corresponding data augmentation strategies are listed. Here ``CONV Aug" means conventional data augmentation, ``TTS Aug" means TTS based data augmentation, ``TTA" means test-time data augmentation, and ``Delta" means we use the original PPG feature and its first and second difference for training.}
  \label{tab:result-dev}
  \centering
  \begin{tabular}{ c|c|l|c|c|c|c|c|c|c|c|c}
  \hline\hline
    \multirow{2}{*}{\textbf{ID}} & \multirow{2}{*}{\textbf{Model}} & \multirow{2}{*}{\textbf{Configuration}} & \multicolumn{9}{c}{\textbf{Accuracy (\%)}} \\\cline{4-12}
      & & & \textbf{US} & \textbf{UK} & \textbf{CHN} & \textbf{IND} & \textbf{JPN} & \textbf{KR} & \textbf{PT} & \textbf{RU} & \textbf{Avg.} \\\hline\hline
       1 & \multirow{8}{*}{TDNN} & FBANK & 36.54 & 90.13 & 54.32 & 92.61 & 51.34 & 35.94 & 73.08 & 49.44 & 60.24 \\
       2 &  & \ \ +CONV Aug & 53.86 & 87.92 & 71.39 & 88.58 & 58.00 & 58.30 & 79.15 & 52.97 & 68.74 \\
       3 &  & \ \ \ \ +TTS Aug & 54.70 & 90.51 & 71.67 & 96.34 & 65.32 & 64.27 & 85.71 & 61.76 & 73.66 \\ 
       4 &  & PPG & 78.89 & 92.28 & 81.20 & 98.71 & 76.01 & 83.61 & 84.47 & 76.79 & 83.74  \\ 
       5 &  & \ \ +CONV Aug & 77.07 & 91.84 & 86.78 & 99.39 & 77.96 & 87.11 & 86.88 & 78.65 & 85.57  \\
       6 &  & \ \ \ \ +TTS Aug & 80.43 & 93.04 & 94.48 & 99.54 & 81.72 & 90.33 & 90.90 & 87.19 & 89.74 \\ 
       7 &  & \ \ \ \ \ \ \ \ +TTA & 80.79 & 92.98 & 94.87 & 99.70 & 83.94 & 88.55 & 91.96 & 89.29 & 90.32 \\
       8 &  & \ \ \ \ \ \ \ \ \ \ +Delta & 81.63 & 92.86 & 94.20 & 99.70 & 84.07 & 87.11 & 93.19 & 87.68 & 90.10 \\\hline 
       9 & \multirow{8}{*}{RES2SETDNN} & FBANK & 50.00 & 82.86 & 45.23 & 85.83 & 46.51 & 32.58 & 70.24 & 48.39 & 57.32  \\
       10 &  & \ \ +CONV Aug & 65.78 & 88.61 & 66.59 & 84.23 & 60.28 & 46.78 & 83.35 & 54.64 & 68.73 \\
       11 &  & \ \ \ \ +TTS Aug & 68.44 & 92.35 & 75.07 & 94.82 & 68.48 & 61.32 & 90.66 & 64.42 & 76.85 \\
       12 &  & PPG & 82.12 & 91.71 & 81.99 & 98.93 & 77.96 & 82.44 & 86.94 & 76.11 & 84.51  \\ 
       13 &  & \ \ +CONV Aug & 81.56 & 92.17 & 89.68 & 99.31 & 79.44 & 88.54 & 88.74 & 83.04 & 87.71  \\
       14 &  & \ \ \ \ +TTS Aug & 79.19 & 93.23 & 94.25 & 99.31 & 81.45 & 90.47 & 93.25 & 87.25 & 89.86 \\ 
       15 &  & \ \ \ \ \ \ +TTA & 80.86 & 93.17 & 94.70 & 99.77 & 84.95 & 86.76 & 93.32 & 87.19 & 90.15 \\
       16 &  & \ \ \ \ \ \ \ \ +Delta & 81.49 & 93.30 & 94.59 & 99.62 & 86.16 & 88.41 & 92.14 & 88.49 & 90.56 \\\hline
       17 & \multicolumn{2}{c|}{challenge baseline} & 60.2 & 93.9 & 67.0 & 97.0 & 73.2 & 55.6 & 85.5 & 75.7 & 76.1 \\\hline
       18 & \multicolumn{2}{c|}{our final system} & 82.68 & 93.36 & 95.37 & 99.77 & 84.88 & 88.96 & 93.69 & 89.85 & 91.13 \\\hline\hline 
  \end{tabular}
\end{table*}

\subsection{Test time augmentation}
\label{TTA}
%%%
%%
% some detailed settings should be placed in the experiment section.
%%
%%%
Test time augmentation is a common trick in image classification tasks to improve the test accuracy~\cite{krizhevsky2012imagenet,simonyan2014very}.
Instead of predicting the label of test image itself, the model takes multiple augmented versions of the test image as input, and the predicting results are then aggregated to give the final result.
In our work, similar test time augmentation for speech data is adopted.
We use the {\it tempo} effect of the SoX tool to do data augmentation by changing the speech rates.
The augmented versions of the test file are then appended to the original test file, and we test our model on the resulting file.

\section{Experiments}
\label{sec:exp}
A detailed comparison of our systems are presented in this section.
Kaldi is used for FBANK feature extraction and PyTorch~\cite{paszke2019pytorch} is utilized to train the neural network models and PPG feature extraction.
All experiment results on the development set are shown in Table~\ref{tab:result-dev}, where 8 accented English are denoted by their corresponding country codes respectively: \textbf{US} (the USA), \textbf{UK} (the United Kingdom), \textbf{CHN} (China), \textbf{IND} (India), \textbf{JPN} (Japan), \textbf{KR} (Korea), \textbf{PT} (Portugal) and \textbf{RU} (Russia).

\subsection{Baseline systems}
\label{baseline}
Our baseline systems are trained on the original 160-hour accent training data, with 40-dimensional FBANK feature. In Table~\ref{tab:result-dev}, the baseline system for TDNN and RES2SETDNN are denoted by ID 1 and 9 respectively.

\subsection{Conventional data augmentation}
We then apply the conventional data augmentation to the accent training data.
First, the speech rate of each utterance in the training set is randomly changed to 0.8$\times$, 0.9$\times$, 1.1$\times$ or 1.2$\times$, which increase the amount of the data to 320 hours.
Then, by augmentation with additive noise and reverberation, we extend the training data to 1600 hours.
The TDNN and RES2SETDNN models trained on the extended training data are shown in Table~\ref{tab:result-dev} with ID 2 and ID 10 respectively. 
Compared with the baseline systems, on average accuracy the conventional data augmentation can give an absolute improvement of 8.50\% and 11.41\% respectively.
Besides, the improvement is consistent across all 8 accents.

\subsection{TTS based data augmentation}
\label{exprtment-tts}
In addition to the conventional data augmentation, we further apply the TTS based data augmentation approach described in Section~\ref{tts-train} on the 1600-hour augmented data to generate 4800-hour synthesized data.

To check the correlation between the synthesized speech and its corresponding accent, we test the synthesized data using system ID 2 and 10 in Table~\ref{tab:result-dev} which are trained on the conventional augmented data.
From Table~\ref{tab:tts-effectiveness}, we can find that the accent identification accuracy on the synthesized data is comparable to that of the development data.

\begin{table}[!htb]
  \caption{Accuracy (\%) on synthesized data}
  \label{tab:tts-effectiveness}
  \centering
  \begin{tabular}{ c||c|c}
  \hline\hline
    \multirow{2}{*}{\textbf{Accent}} & \multicolumn{2}{c}{\textbf{System ID}}  \\
    \cline{2-3} & \textbf{2} & \textbf{10}\\\hline\hline
       US & 59.64 & 60.50 \\
       UK & 90.73 & 86.04 \\ 
       CHN  & 70.93 & 64.08 \\
       IND & 67.13 & 70.70 \\ 
       JPN  & 63.42 & 53.73 \\
       KR & 67.48 & 66.91 \\ 
       PT  & 86.44 & 81.77 \\
       RU & 46.27 & 45.04 \\ 
       Avg. & 69.08 & 66.16 \\\hline\hline
  \end{tabular}
\end{table}

\begin{table*}[!thb]
\caption{Final submission results on the accent identification evaluation set of the top 4 teams in the rank list}
  \label{tab:final}
  \centering
  \begin{tabular}{ c|c|c|c|c|c|c|c|c|c|c}
  \hline\hline
    \multirow{2}{*}{\textbf{Ranking}} & \multirow{2}{*}{\textbf{Team}} &  \multicolumn{9}{c}{\textbf{Accuracy (\%)}} \\\cline{3-11}
      & & \textbf{US} & \textbf{UK} & \textbf{CHN} & \textbf{IND} & \textbf{JPN} & \textbf{KR} & \textbf{PT} & \textbf{RU} & \textbf{Avg.} \\\hline\hline
       1 & S2 (ours) & 65.64 & 94.77 & 87.6 & 97.11 & 81.49 & 83.43 & 79.66 & 85.25 & 83.63 \\
       2 & E2 & 52.55 & 93.11 & 60.65 & 90.41 & 68.43 & 79.12 & 76.47 & 65.83 & 72.39 \\
       3 & Z2 & 37.64 & 90.87 & 72.79 & 92.32 & 61.65 & 78.90 & 69.93 & 63.19 & 69.63 \\ 
       4 & F & 33.55 & 89.60 & 62.75 & 89.37 & 69.73 & 83.09 & 77.46 & 62.26 & 69.59  \\ 
       * & challenge baseline & 40.2 & 89.4 & 57.4 & 88.4 & 62.4 & 53.8 & 63.6 & 63.8 & 64.9  \\\hline\hline
  \end{tabular}
\end{table*}

We train system ID 3 and 11 with the combined 6400-hour data.
As shown in Table~\ref{tab:result-dev}, the average accuracy is largely improved again, also, the improvement across the 8 accents is consistent.
This verifies the effectiveness of our proposed TTS based data augmentation.

\subsection{PPG features versus FBANK features}
\label{exprtment-ppg}
In this section, we compare the systems trained with PPG features or FBANK features.
% The PPG features of the accent training data are extracted with the SI-ASR model trained in \cite{tan2020aispeech}.
The SI-ASR model for PPG feature extraction is prepared with the same setting in~\cite{tan2020aispeech}.
Similar to the training data augmentation schemes for systems trained with FBANK features, we train the TDNN and RES2SETDNN models on the original data and two sets of augmented data respectively.
The system trained on PPG features are denoted by ID 4, 5, 6 and 12, 13, 14 in Figure~\ref{tab:result-dev}.

According to the reported numbers of the average accuracy in Table~\ref{tab:result-dev}, the system ID 4 (or 12) beats the system ID 1 (or 9) by a large margin, though they are trained on the original data only.
This suggests that, with the speaker independent property, PPG features make the discrimination of accent much easier.
In addition, comparing system ID 5, 6 (or 13, 14) to ID 4 (or 12), we can find the system performance can be improved when applying data augmentation.

In this work, the SI-ASR system used as extractor for PPG features is trained over all available data (accented+librispeech). When only the accented data are used for training the SI-ASR system, PPG features achieves about 3\% performance improvement compared to the FBANK features on average accuracy, which is small then using all data.

\subsection{Test-time data augmentation}
Table~\ref{tab:result-dev} shows, with the test-time data argumentation described in Section~\ref{TTA}, the performance has a slight lift from system ID 6 to 7 (or ID 9 to 10) .

\subsection{Delta PPG features}
\label{exp-tta}

We train systems with 120-dimensional delta PPG features (the original feature and its first and second difference) on the combined 6400 hours training data.
We find the performance has a small drop on the TDNN model (system ID 7 and 8), but rises a little on the RES2SETDNN model (system ID 15 and 16).

\subsection{Final system}

In our final system, following the design in Section~\ref{hm}, we first initialize the embedding extraction part with four systems (ID 7, 8, 15 and 16) in Table~\ref{tab:result-dev} and fix the parameters.
Then we train the fully connected layer for accent classification.
The final prediction is given by the 8-way classification hierarchical multi-embedding joint model on the test time augmented wave file. As shown in Table~\ref{tab:result-dev}, our final system (ID 18) significantly outperforms the challenge baseline system (ID 17), reaching a remarkable 15.03\% accuracy improvement on the development set.

On the challenge evaluation data, Table~\ref{tab:final} shows the rankings of the leading submissions for this challenge. Our system ranks first in the challenge with an average accuracy of 83.63\% on the evaluation set, ahead of the second team by 11.23\%.
Moreover, when comparing the system performance shown in Table~\ref{tab:result-dev}, our final system achieves much smaller performance gap between the development data and the evaluation data than the challenge baseline.

\begin{figure}[!htb]
\centering{\includegraphics[scale=0.5]{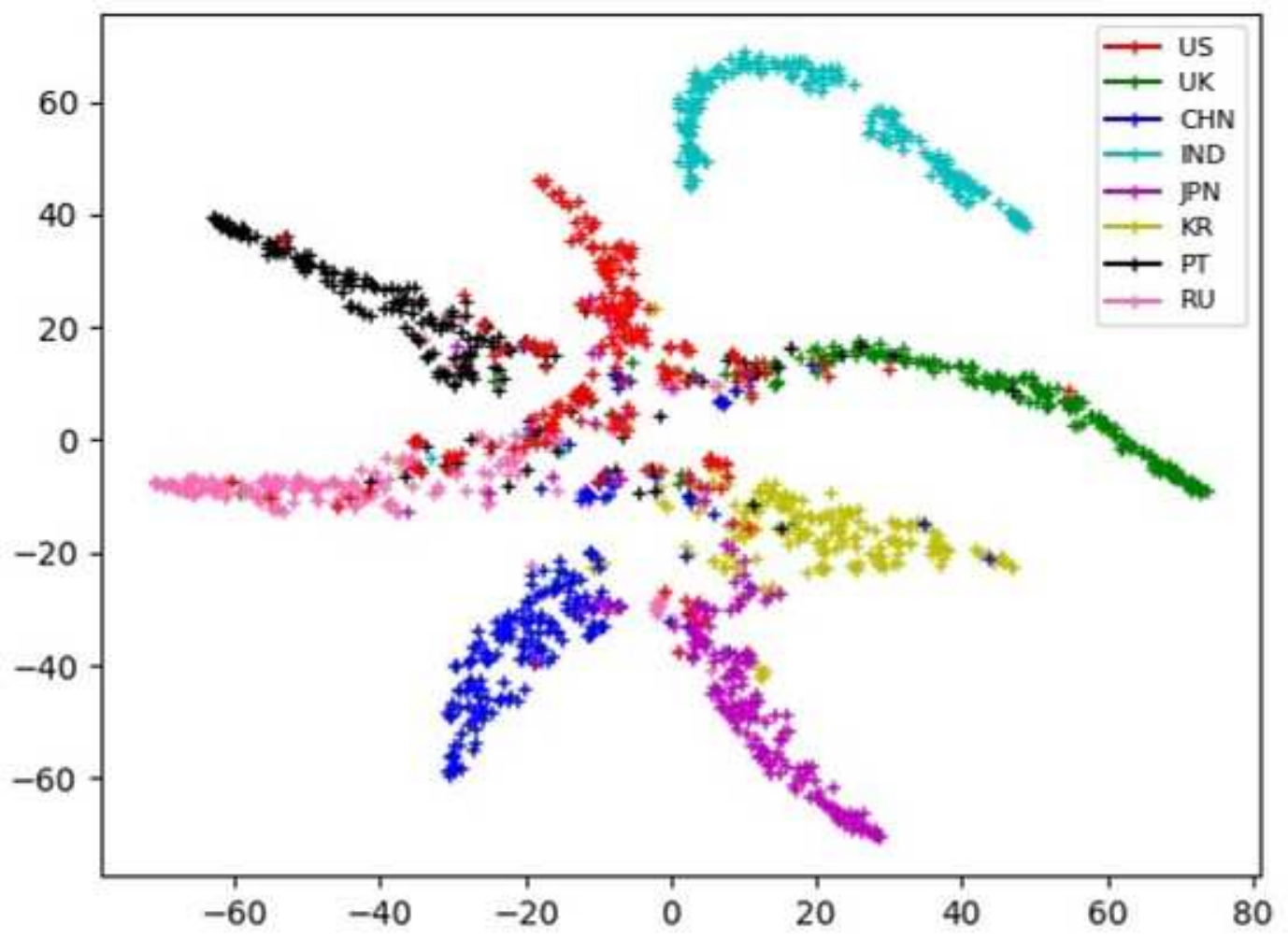}}
\caption{Accent embeddings of 8 accents in test set. 200 accent embeddings for every accent are chosen in the test set.}
\label{fig:t-SNE}
\end{figure}

To better understand our system performance on each accent, we visualize the accent embeddings of the test data using t-SNE~\cite{2008Visualizing}. As shown in Figure~\ref{fig:t-SNE}, the cluster of the {\bf IND} utterances is compact and far from the other clusters, while the cluster of the {\bf US} utterances has many overlaps with the other clusters. This partially explains the differences in identification accuracy.

\section{Conclusions}
\label{sec:conclusion}

In this paper, we describe our submitted system for the Interspeech-2020 Accented English Speech Recognition Challenge (AESRC). Several novel approaches are developed to improve the robustness of our accent identification system. To the best of our knowledge, it is the first time phone posteriorgram feature has been introduced to accent classification, which brings an improvement of more than 15\% compared to the regular FBANK feature. To train a robust system from such limited data, we adopt TTS based data augmentation to synthesize additional accented training data, improving the system performance by 2.15\%$\sim$4.17\%. Test-time data augmentation and hierarchical multi-embedding joint model training are employed for further boosting the system performance.
Based on these approaches, our final system achieves an average accent identification accuracy of 83.63\% on the AESRC evaluation set, ranking first among all the participants. We find that the evaluation set is more challenging than the development set, which leads to more than 7\% performance degradation. In the future, we will focus on analyzing and narrowing this performance gap.
 
\vfill\pagebreak

% \section{REFERENCES}
% \label{sec:refs}

% List and number all bibliographical references at the end of the
% paper. The references can be numbered in alphabetic order or in
% order of appearance in the document. When referring to them in
% the text, type the corresponding reference number in square
% brackets as shown at the end of this sentence \cite{C2}. An
% additional final page (the fifth page, in most cases) is
% allowed, but must contain only references to the prior
% literature.

% References should be produced using the bibtex program from suitable
% BiBTeX files (here: strings, refs, manuals). The IEEEbib.bst bibliography
% style file from IEEE produces unsorted bibliography list.
% -------------------------------------------------------------------------
\bibliographystyle{IEEEbib}
\bibliography{refs}

\end{document}